\documentclass[sigconf]{acmart}

\renewcommand\footnotetextcopyrightpermission[1]{} 
\usepackage{amsfonts}
\usepackage{amsmath}
\usepackage{amsthm}
\usepackage{array}
\usepackage{bm}
\usepackage{booktabs}
\usepackage{bxtexlogo}
\usepackage{caption}
\usepackage{float}
\usepackage{geometry}
\usepackage{graphicx}
\graphicspath{{figs/}}
\usepackage[utf8]{inputenc}
\usepackage{makecell}
\usepackage{mathtools}
\usepackage{multirow}
\usepackage{natbib}
\setcitestyle{authoryear,numbers,aysep={}}
\usepackage{soul}
\usepackage{stfloats}
\usepackage{subcaption}
\usepackage[switch]{lineno}
\def\innerprod<#1>{\langle #1 \rangle}

\begin{document}
\title{Towards Realistic and Interpretable Market Simulations: \\
Factorizing Financial Power Law using Optimal Transport}

\author{Ryuji Hashimoto}
\affiliation{%
  \institution{The University of Tokyo}
  \city{Tokyo}
  \country{Japan}}
\email{hashimoto-ryuji419@g.ecc.u-tokyo.ac.jp}
\orcid{0009-0008-0042-1477}

\author{Kiyoshi Izumi}
\affiliation{%
  \institution{The University of Tokyo}
  \city{Tokyo}
  \country{Japan}}
\email{izumi@sys.t.u-tokyo.ac.jp}

\renewcommand{\shortauthors}{Hashimoto et al.}

\begin{abstract}
We investigate the mechanisms behind the power-law distribution of stock returns using artificial market simulations. While traditional financial theory assumes Gaussian price fluctuations, empirical studies consistently show that the tails of return distributions follow a power law. Previous research has proposed hypotheses for this phenomenon---some attributing it to investor behavior, others to institutional demand imbalances. However, these factors have rarely been modeled together to assess their individual and joint contributions. The complexity of real financial markets complicates the isolation of the contribution of a single component using existing data. To address this, we construct artificial markets and conduct controlled experiments using optimal transport (OT) as a quantitative similarity measure. Our proposed framework incrementally introduces behavioral components into the agent models, allowing us to compare each simulation output with empirical data via OT distances. The results highlight that informational effect of prices plays a dominant role in reproducing power-law behavior and that multiple components interact synergistically to amplify this effect.
\end{abstract}


\maketitle

\section{Introduction}
In financial markets, certain phenomena contradict the assumptions of traditional finance. For example, theories of financial engineering assume that the distribution of stock returns is Gaussian, although empirical studies have found a power law distribution~\citep{power_law_return}. This discrepancy suggests that the risk of extreme price movements is much higher in practice than that predicted by the Gaussian distribution. Furthermore, the exponent of this power law consistently approximates 3 (universal cubic law ~\citep{cubic_law}). The cubic law is a widely observed financial stylized fact across various markets worldwide~\citep{empirical_power_law_return1,empirical_power_law_return2}.

The cubic law mechanism has attracted significant interest from researchers, leading to various theoretical hypotheses about its underlying components~\citep{financial_power_laws_survey}. \citet{cubic_law} modeled the cubic law combined with the distribution of large investors' sizes. \citet{herd_behavior} focused on herd behavior. \citet{info_effect} examined the informational effect of prices. However, neither the individual contributions of these components nor the combination effect of multiple components have been evaluated.

\begin{figure}[t]
\centering
\includegraphics[bb=0 0 836 510, width=8.5cm]{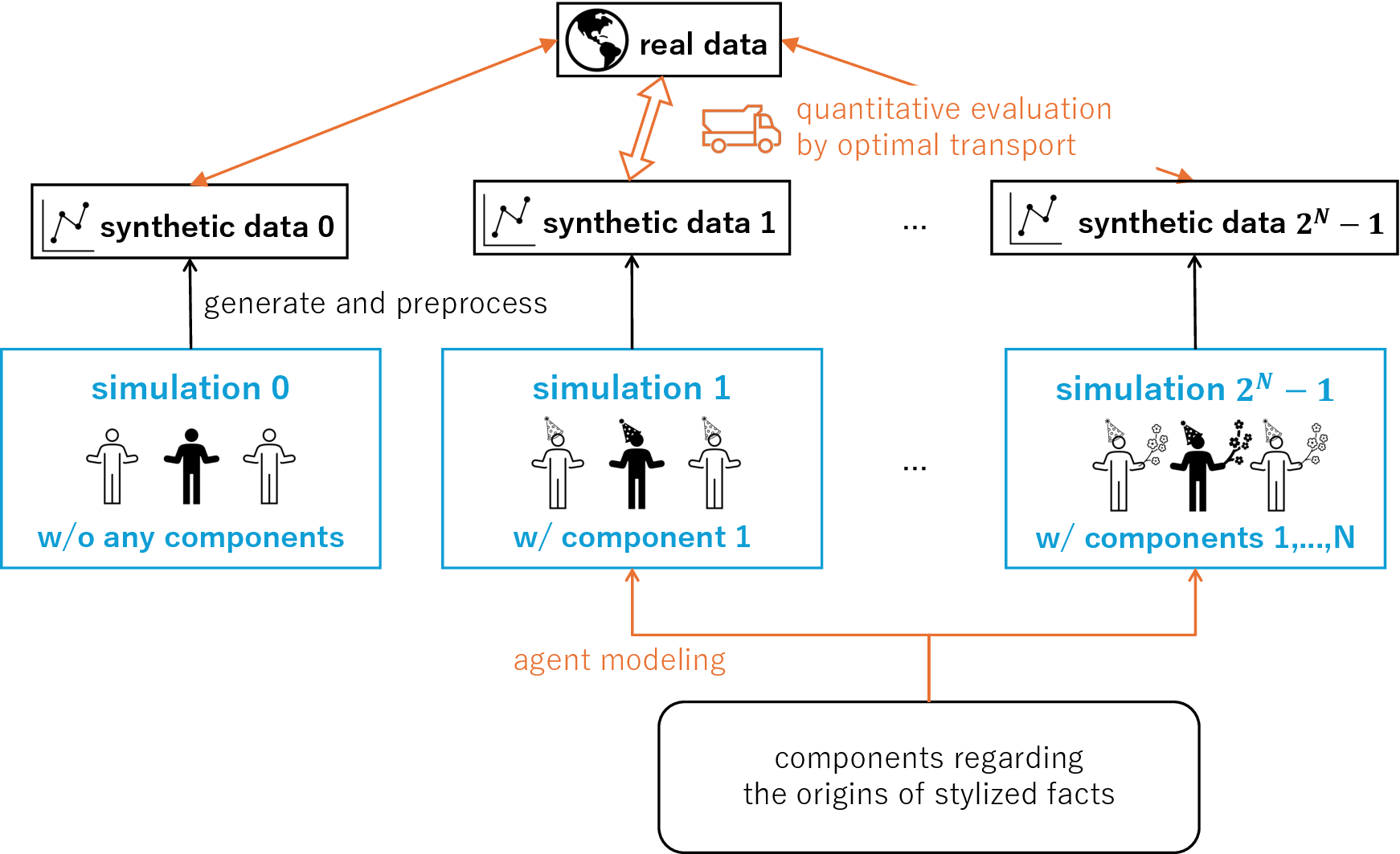}
\caption{Our proposed pipeline for hypothesis evaluation cycle of agent models. Each component, hypothesized to contribute to the emergence of a given stylized fact, is iteratively introduced into the agent model, generating multiple simulation scenarios that range from a baseline model (without any components) to model with all candidate components. The resulting synthetic data from each simulation is assessed using OT against real market data.}
\label{Fig:proposed_pipeline}
\end{figure}

Artificial markets, a form of multi-agent simulation, offer a constructive approach to understanding financial phenomena---such as the cubic law---from a microscopic perspective. To replicate macro-level patterns, researchers design heterogeneous agents whose interactions generate emergent market behavior. Although artificial markets are expected to reveal the underlying mechanisms of financial systems~\citep{am2,am1,am3,am4}, a concrete methodological framework remains elusive. One major limitation is the lack of a quantitative criterion to compare simulation results. Most studies rely on descriptive statistics related to stylized facts~\citep{get_real, stylized_facts} to validate their simulations. However, these statistics serve only as qualitative benchmarks and cannot be used as quantitative {\em lower-is-better} metrics to assess the similarity between synthetic and real data.

To factorize the cubic law and quantitatively interpret the realism of artificial market simulations, we propose a method for hypothesis evaluation cycles for agent models based on optimal transport (OT)-based evaluation. As described in Figure~\ref{Fig:proposed_pipeline}, the proposed pipeline first conducts simulations with and without reflecting the multiple components to be verified in the agent model and then compares the results of each simulation with real data using OT. Artificial markets enable us to conduct such controlled experiments and determine the extent to which each component contributes to the formation of the stylized fact and the existence of interactions among the components. The proposed evaluation method measures the OT distance between point clouds constructed from real and synthetic data, enabling the assessment of the realism of synthetic data from arbitrary perspectives.

In our experiment, we first choose three components as candidates for the origins of the cubic law in the high-frequency domain: the power law distribution of large investors' sizes~\citep{cubic_law}, the informational effect of prices~\citep{info_effect}, and herd behavior~\citep{herd_behavior}. We then reflect these components in our agent model and conducted $8(=2^3)$ different simulations with and without each component. Each synthetic data extracted from these simulation results is evaluated by OT distance from real data and Hill index~\citep{hill}. Our experiment reveals that each of the above-mentioned candidate components bring the tail distribution of synthetic price fluctuations closer to that of real data. Additionally, the experimental results suggest that each of the above candidates acts in combination, centering on the chartist trader, that is, the informational effect of prices.

To summarize, our contributions are as follows:
\begin{itemize}
\item We propose methods for 1) factorizing financial stylized facts through artificial market simulations with hypotheses evaluation cycles of agent models, and 2) calibrating simulations based on quantitative evaluation of financial synthetic data using OT. 
\item We demonstrate these methods by applying them to universal cubic law. We select the power law of demand size, the informational effect of prices, and herd behavior as candidate components for the cubic law and incorporat each component into the agent model.
\item Our experiment illustrate that 1) the informational effect of prices plays a dominant role in generating the cubic law, and that 2) a synergistic effect exists between the power law of demand size and informational effect of prices.
\end{itemize}

\section{Related Work\label{Sec:related_work}}

\subsection{Universal Cubic Law}
Empirical studies motivated by the increasing availability of high-frequency data have revealed that the tail distribution of stock returns exhibits power law behavior~\citep{power_law_return}. Let $r_t$ denote the stochastic variable representing the logarithmic stock return over a time interval $\Delta t$, standardized to have a mean of $0$ and a standard deviation of $1$. The distribution of the absolute return $|r_t|$ asymptotically follows:
{\small
\begin{eqnarray}
Pr[x\leq |r_t|]\sim x^{-\zeta_r}\label{Eq:def_power_law}
\end{eqnarray}
}
Here, $\zeta_r$ denotes the power law exponent. Equation~(\ref{Eq:def_power_law}) suggests that the tail distribution of stock returns decays more slowly than the exponential function $e^{-x}$, indicating that tail risk cannot be ignored. Tail risk refers to the risk beyond the confidence interval that cannot be fully captured by risk metrics, such as Value at Risk (VaR), which are calculated based on quartiles of a Gaussian distribution~\citep{var_pitfalls}.

Empirical studies have estimated the power law exponent, $\zeta_r$, using a variety of real data. For instance, \citet{empirical_power_law_return2} reported a Hill index~\citep{hill} of $3.35$ using one-minute bar price series of DAX from the late 20th century. Let $r_{(N)},...,r_{(1)}$ denote the descending order statistics of a sample of absolute log-returns $|r_1|,...,|r_N|$ of size $N$, the Hill index, $\hat{\zeta}_r$ is then given by:
{\small
\begin{eqnarray}
\tilde{r}_{(k)}&=&\log \frac{r_{(N-k+1)}}{r_{(N-K)}},~ k=1,...,K\label{Eq:r_tilde}\\
\hat{\zeta}_r&=&K/\sum_{k=1}^K\tilde{r}_{(k)}\label{Eq:hill}
\end{eqnarray}}
$K$ is a cutoff threshold. In practice, the upper $5\%$ of the sample set is used to determine the value of $K$. Estimates of the power-law exponent, $\hat{\zeta}_r$, using data from other markets and with different sampling frequencies, as reported in \citep{empirical_power_law_return1,empirical_power_law_return3}, have consistently been around $3$. This property has been termed the {\em universal cubic law} by \citet{cubic_law}.

Various theoretical hypotheses have been proposed regarding the emergence of the cubic law. Generally, two major principles underlie the mechanism of power laws. First, the power law emerges because of a multiplicative stochastic process. If institutional investors' wealth growth follows a multiplicative process with a {\em rich get richer} property, the distribution of demand sizes among market participants will exhibit a fat tail, leading to significant price impacts from large orders. \citet{rich_get_richer_return} analyzed individual investors' asset positions and found that portfolios with higher valuations had higher risk-adjusted returns. \citet{cubic_law} demonstrated that the distribution of returns follows a power law because the size of institutional investors, and hence the demand size, follows a power law. Second, from a multi-agent system's viewpoint, if agents act collectively akin to a spin system, an order with a power law appears under certain conditions. When investors gradually synchronize their behaviors in response to price fluctuations, price fluctuations amplify. \citet{info_effect} proposed a model in which the investor outlook depends on market prices by assuming information asymmetry. Specifically, they proved the power law in stock returns through herd behavior driven by price changes. Market price increases attract some investors to buy the stock because they believe that other investors possess private information indicating that the stock price will continue to rise. Alternatively, connections through social networks, independent of prices, can also induce correlations in investment behavior. \citet{herd_behavior} formulated a multi-agent model in which market participants changed their outlook to that of other participants. They claimed that large price fluctuations occur because of the switching process of majority opinion induced by the herding mechanism.

While the components in these hypotheses likely play roles in real markets to a certain degree, no existing research has examined 1) the magnitude of influence of each component on the cubic law and 2) the interactions among these components.

\subsection{Artificial Market Simulation}
\subsubsection{FCNAgent}
An artificial market is a virtual financial market simulated on a computer. Research on artificial markets aims to understand macro-level phenomena by modeling individual investors as stochastic agents. \citet{mas_essential} demonstrated that the unique phenomena observed in financial markets cannot be replicated without inter-agent interactions, thus highlighting the necessity of multi-agent simulations. Various types of social experiments can be conducted using heterogeneous agents that imitate real investors' trades according to the regulations of real financial markets. The agent model is designed based on the ordering patterns found in empirical studies. For example, \citet{fcn1} proposed FCNAgent model that predicts future stock returns $\hat{r}_{t+\tau^j}^j$ and price $\hat{p}_{t+\tau^j}^j$ given simulation end time $T_{sim}$, the current time steps $t\in\{1,...,T_{sim}\}$, market price $p_t$, and fundamental price $p_t^f$ as:
\begin{eqnarray}
\hat{r}_{t+\tau^j}&=& \frac{1}{w^{j,f}+w^{j,c}+w^{j,n}}\nonumber\\
&& \left(\frac{w^{j,f}}{\tau^{j,f}}\log\frac{p_t^f}{p_t}+\frac{w^{j,c}}{\tau^j}\log\frac{p_t}{p_{t-\tau^j}}+w^{j,n}\epsilon_t\right)\label{Eq:fcn_r}\\
\hat{p}_{t+\tau^j}^j&=& p_t\exp(\tau^j\hat{r}_{t+\tau^j}^j)\label{Eq:fcn_p}
\end{eqnarray}
Let $n_a$ denote the number of agents. $j\in\{1,...,n_a\}$ represents the number individually assigned to each agent, $\tau^j$ denotes the time window size of agent $j$. $w^{j,f},w^{j,c},w^{j,n}$ are the coefficients corresponding to the three components described below, randomly determined for each agent such that $w^{j,f}\sim Ex(\lambda^f)$, $w^{j,c}\sim Ex(\lambda^c)$, $w^{j,n}\sim Ex(\lambda^n)$. Here, $Ex(\lambda)$ indicates an exponential distribution with expected value $\lambda$. The first term inside the parentheses in Equation~(\ref{Eq:fcn_r}) $\frac{w^{j,f}}{\tau^{j,f}}\log\frac{p_t^f}{p_t}$ is called the fundamental trader component, implying that agent $j$ predicts that market price $p_t$ will reverse to fundamental price $p_t^f$ in $\tau^{j,f}$ steps. The second term $\frac{w^{j,c}}{\tau^j}\log\frac{p_t}{p_{t-\tau^j}}$ is called the chartist trader component, modeling a typical ordering pattern, the positive feedback trader~\citep{feedback,noise}, which reflects past price movements in return predictions. The last term $w^{j,n}\epsilon_t$ is called the noise trader component. $\epsilon_t$ is Gaussian noise with a mean $0$ and variance $(\sigma^n)^2$. FCNAgent was named after its three components, fundamental, chartist, and noise traders, which together form its return prediction model.

\citet{fcn2} proposed the order decision of the FCNAgent based on price prediction $\hat{p}_{t+\tau^j}^j$, holding cash amount $c_t^j$, holding stock position $w_t^j$, and the risk-aversion term $\alpha^j$ by assuming the constant absolute risk aversion (CARA) utility function $\mathcal{U}_t^j$:
\begin{eqnarray}
\mathcal{U}_t^j=-\exp\left\{-\alpha^j(w_t^jp_t+c_t^j)\right\}\label{Eq:cara}
\end{eqnarray}
Here, $\forall j~ c_0^j\sim U(0,c_{max}),~ w_0^j\sim U(0,w_{max})$. $U(0,c_{max})$ indicates uniform distribution with range $[0,c_{max}]$. The agent decides whether to buy or sell the stock, limit price, and order volume to maximize their expected future utility $\mathbb{E}_t[\mathcal{U}_{t+\tau^j}^j]$, where $\mathbb{E}_t[\cdot]$ stands for the expected mean conditional to the information available at time $t$.

Although the chartist trader can be interpreted as representing the informational effect of prices~\citep{info_effect}, FCNAgent lacks any other component in the hypotheses referred to in the previous subsection.

\subsection{Evaluation of Financial Synthetic Data}\label{Sec:evaluation_abms_challenges}
In recent years, interest in financial synthetic data generation has grown ~\citep{financial_synthetic_data_generation1}. Considered a subset of financial synthetic data generation, artificial markets face the challenge of evaluating their realism. Evaluation methods for agent-oriented financial synthetic data can be categorized into three types. The first is evaluation based on stylized facts. Stylized facts~\citep{stylized_facts} are empirical phenomena observed across a wide variety of markets and periods. \citet{get_real} proposed {\em realism metrics} of synthetic data using stylized facts related to stock returns, volumes, and order flow. \citet{stylized_facts_distance} proposed a calibration method for artificial markets that minimizes a weighted sum of {\em stylized facts distances}, such as the difference in the stock return distributions between real and synthetic data. However, the features that can be used to evaluate realism using these methods are limited to those for which stylized facts have been established. The second is the method of simulated moments (MSM). MSM~\citep{msm1,msm2} is a method of structural estimation that aggregates the errors of multiple moment statistics defined for both real and synthetic data. \citet{msm_rl} proposed a calibration method for economic simulation models using an objective function formulated by MSM and reinforcement learning-based exploration. However, the effectiveness of MSM depends on the assumed distribution. MSM is not suitable when the underlying distribution of the real data is assumed not to have moments. As a whole, the above-mentioned methods that rely on summary statistics risk losing crucial information about the underlying order in financial data. The third is a method to evaluate the predictive accuracy of agent models as models of individual investor behavior~\citep{lstm_hftmm,calibration_w_nde}. \citet{lstm_hftmm} evaluated their deep learning-based agent model by computing Kullback-Leibler (KL) divergence between the outputs of their model and real trading behavior data. This method requires historical data at the individual investor level, which is not freely available. As shown above, a quantitative evaluation method for the realism of arbitrary aspects of synthetic data has not yet been established.

\section{Experiment\label{Sec:experiment}}

\begin{table*}[tbp]
\centering
\caption{Hypotheses on the origins of the cubic law and corresponding components in our agent model. Simulation No. column shows the correspondence between the simulation numbers and the inclusion of each component for the simulation.}
\begin{tabular*}{14.5cm}{llccccccccc}
\midrule
\multirow{2}{*}{Candidate component}&\multirow{2}{*}{Component in our agent model}&\multicolumn{8}{l}{Simulation No.}\\
                                 &                                                     & 0 & 1            & 2            & 3            & 4            & 5            & 6            & 7 \\
\midrule
The power law of demand size~\citep{cubic_law} & Pareto cash amount. Equation~(\ref{Eq:pareto_cash_amount}) &   & $\checkmark$ &              &              & $\checkmark$ & $\checkmark$ &              & $\checkmark$\\
Informational effect of prices~\citep{info_effect}  & Chartist trader. $0<\lambda^c$                      &   &              & $\checkmark$ &              & $\checkmark$ &              & $\checkmark$ & $\checkmark$\\
Herd behavior~\citep{herd_behavior}     & Mood-aware trader. $0<\lambda^m$                    &   &              &              & $\checkmark$ &              & $\checkmark$ & $\checkmark$ & $\checkmark$\\
\midrule
\end{tabular*}
\label{Tab:hypotheses_agents}
\end{table*}

\begin{figure}[tbp]
\centering
\includegraphics[bb=0 0 809 503, width=8.5cm]{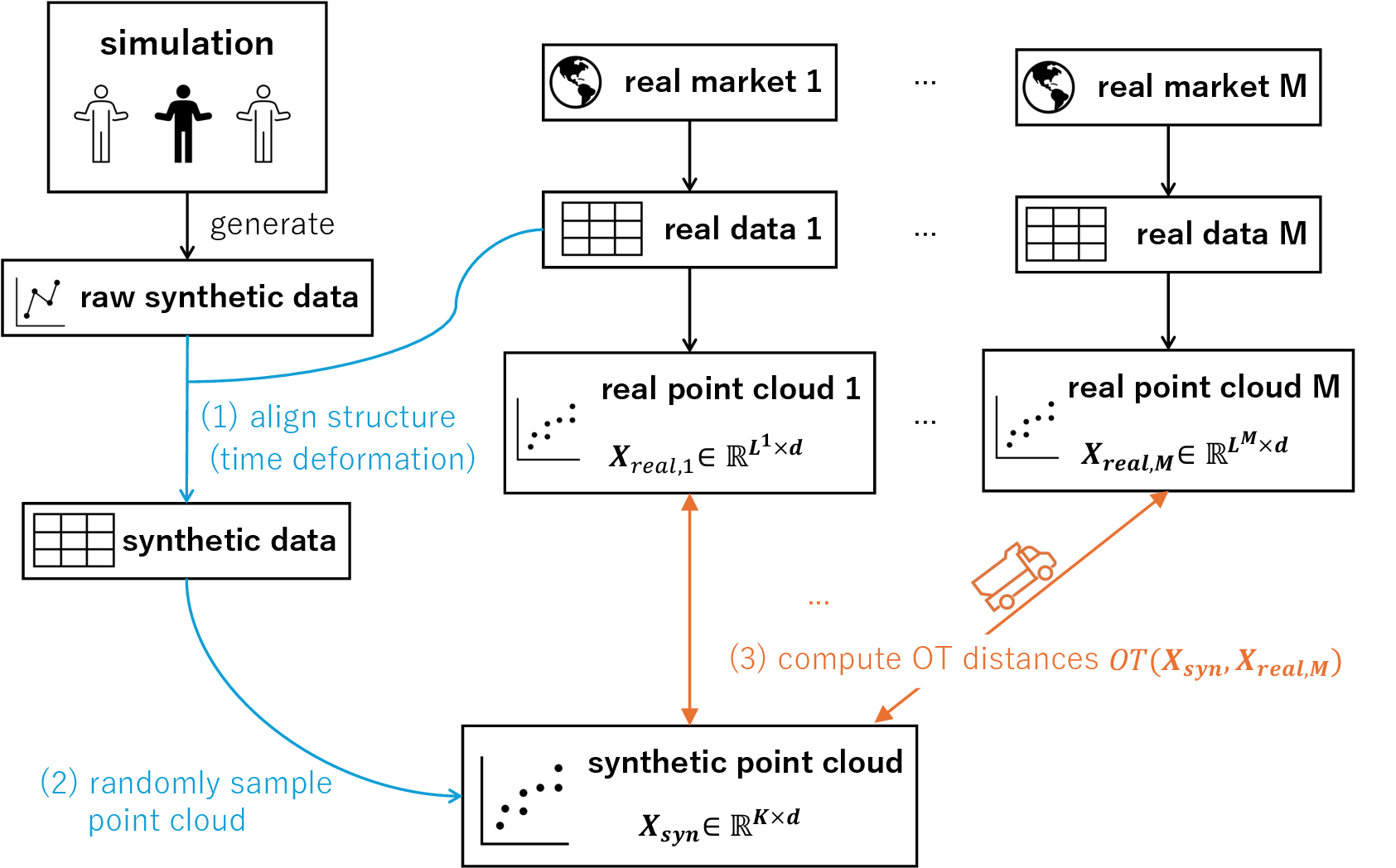}
\caption{Our proposed pipeline of an evaluation of an artificial market simulation. We evaluate the realism of the simulation with the following procedure: 1)extract structured data, 2)construct pre-defined point cloud, 3)measure OT distance between real and synthetic point clouds.}
\label{Fig:ot_evaluation}
\end{figure}

In our experiments, we incorporated three candidate components for the origins of the cubic law in the high-frequency domain into the agent model. Subsequently, we conducted $8 (=2^3)$ types of simulations, varying whether each of the three components was included in the agent model, and evaluate the results against real data through the pipeline illustrated in Figure~\ref{Fig:ot_evaluation}.

\subsection{Agent Model}

We used the agent model that modifies the FCNAgent to reflect hypotheses on the origins of the cubic law, power law of demand size, informational effect of prices, and herd behavior. As mentioned above, the FCNAgent already has a component that expresses the informational effect of prices as a chartist trader. Hence, we added the remaining components, that is, the power law of demand size and herd behavior to the FCNAgent. In order to add the power law of demand size to the FCNAgent, we changed the distribution of the initial cash amount $c_0^j$ of agent $j\in\{1,...,n\}$ from a uniform distribution to a Pareto distribution:

\begin{eqnarray}
c_0^j\sim Pareto(c_{min},\beta)\label{Eq:pareto_cash_amount}
\end{eqnarray}
Here, $Pareto(c_{min},\beta)$ denotes a Pareto distribution with scale and shape parameter $c_{min},\beta\in\mathbb{R}_+$.

Also, to model herd behavior,  we modified the FCNAgent's return prediction (Equation~(\ref{Eq:fcn_r})) as follows:
{\small
\begin{align}
\MoveEqLeft \hat{r}_{t+\tau^j}=\frac{1}{w^{j,f}+w^{j,c}+w^{j,m}+w^{j,n}}\nonumber\\
\MoveEqLeft \left(\frac{w^{j,f}}{\tau^{j,f}}\log\frac{p_t^f}{p_t}+\frac{w^{j,c}}{\tau^j}\log\frac{p_t}{p_{t-\tau^j}}+w^{j,m}(I_{M_t^j}(O)-I_{M_t^j}(P))+w^{j,n}\epsilon_t\right)\label{Eq:fcmn_r}
\end{align}
}
$M_t^j$ is a random variable called mood that takes two states: optimistic ($O$) and pessimistic ($P$). $I_{M_t^j}:\{O, P\}\mapsto\{0,1\}$ is an indicator function that returns $1$ if the argument matches the current mood $M_t^j$, and $0$ otherwise. We call the third term inside the parentheses in Equation~(\ref{Eq:fcmn_r}) $w^{j,m}(I_{M_t^j}(O)-I_{M_t^j}(P))$ a mood-aware trader component, implying that the agent overestimates their return prediction by $w^{j,m}$ if they are in an optimistic state $O$ and underestimates it if they are in a pessimistic state $P$. $w^{j,m}$ is randomly determined for each agent such that $w^{j,m}\sim Ex(\lambda^m)$. The agent changes their mood according to the following equation:
{\small
\begin{eqnarray}
Pr[M_{t+1}^j=O\mid M_t^j=P]=\frac{\nu n_t^O}{n},~Pr[M_{t+1}^j=P\mid M_t^j=O]=\frac{\nu n_t^P}{n}\label{Eq:mood_transition}
\end{eqnarray}}
where $n_t^O$ denotes the number of optimistic agents at step $t$, $n_t^P$ signifies the number of pessimistic agents, and $\nu$ is an exogenous hyper-parameter. Therefore, $n_a=n_t^O+n_t^P$ holds for all $t\in\{1,...,T_{sim}\}$. We defined optimists rate as $r_t^O=n_t^O/n_a$. The mood transition rule of Equations~(\ref{Eq:mood_transition}) was proposed in ~\citet{herd_behavior}, modeling herd behavior, in which the moods of agents tend to switch with probabilities reflecting the influence of the majority opinion. 

Table~\ref{Tab:hypotheses_agents} summarizes the hypotheses regarding the origins of the cubic law and the corresponding components in our agent model. In our experiment, we conducted $8(=2^3)$ different simulations with and without each component.

\subsection{Evaluation of Synthetic Data with OT}
We evaluated the realism of the simulation results using the framework depicted in Figure~\ref{Fig:ot_evaluation}. We constructed point clouds by sampling points from data collected from both real and artificial markets. In this experiment, we defined a point cloud $X\in\mathbb{R}^{K\times d}$ using samples of log returns, denoted as $r_1,...,r_N$, as follows.
\begin{align}
X={}\top\!\begin{pmatrix}\tilde{r}_{(1)} & \ldots & \tilde{r}_{(K)}\end{pmatrix}
\end{align}
where $d\in\mathbb{N}$ was the dimension of a point. In this experiment, a point was defined as the scalar ($d=1$), i.e. the logarithmic ratio of $k$ and $K$-th largest absolute log-return $\tilde{r}_{(k)}$. This design allowed us to focus on the shape of the tail, not where the tail begins. We denoted the point cloud obtained from the artificial market as $X_{syn}$, and those obtained from the real markets as $X_{real,m},~ m\in\{1,...,M\}$. $M$ denotes the number of real markets. Also, we denoted the $k$-th row element of the point cloud $X$ as $X^k=\begin{pmatrix}\tilde{r}_{(k)}\end{pmatrix}$. We defined the OT distance between the point clouds $X_{syn}$ and $X_{real,m}$ as $OT(X_{syn},X_{real,m})$, which is given by the following equation:

\begin{align}
\begin{split}
\MoveEqLeft OT(X_{syn},X_{real,m})=\min_{P\in\mathbb{R}^{K\times L}}\sum_{k=1}^K\sum_{l=1}^L\|X_{real,m}^k-X_{syn}^l\|_2^2P^{k,l}\\
& s.t.~~ \forall k,l~ 0\leq P^{k,l},~ \sum_{k=1}^KP^{k,l}=\frac{1}{L},~ \sum_{l=1}^LP^{k,l}=\frac{1}{K}
\end{split}\label{Eq:ot_distance}
\end{align}
Here, $P^{k,l}$ denotes the $k$-th row and $l$-th column element of the matrix $P\in\mathbb{R}_+^{K\times L}$. We denote the size of each point cloud for real and synthetic point cloud $K,L$. Equation~(\ref{Eq:ot_distance}) is a type of OT problem~\citep{ot, ot2}. By computing the OT distance between each pair of point clouds $OT(X_{syn},X_{real,m}),~m=1,...,M$, we could measure how closely the shape and structure of extreme stock returns in the simulation matched those in actual market data. In our experiment, we evaluated the realism of the simulation by the average OT distance, calculated as $\bar{OT}=\frac{1}{M}\sum_{m=1}^MOT(X_{syn},X_{real,m})$.

Compared to conventional methods discussed in Section~\ref{Sec:evaluation_abms_challenges}, the proposed approach offers three distinct advantages. First, it preserves the positional information of each sample as a point, enabling the comparison of real and synthetic data while maintaining the inherent geometric structure of the data. Second, unlike KL divergence, which often require overlapping supports and can fail in high-dimensional or sparse settings, OT defines a distance even between non-overlapping distributions. This makes it a more robust measure for assessing distributional discrepancies in complex data spaces. Third, the flexibility of defining each data point in an arbitrary feature space allows our method to evaluate synthetic data from multiple perspectives, without relying solely on predefined stylized facts. This enables researchers to tailor the evaluation to specific aspects of interest---such as local structure, trend, or volatility---by embedding data into an appropriate representation space. Together, these properties allow our framework to capture fine-grained mismatches in distributional shape and local structure, thereby offering a more robust and informative assessment of synthetic data realism.

\subsection{Details of Experimental Settings}
We conducted simulations under $8$ different settings, numbered from $0$ to $7$. The correspondence between the simulation number and the components incorporated into the agent model for each simulation is summarized in Table~\ref{Tab:hypotheses_agents}. Common settings for simulations $0$ to $7$ are as follows.

\begin{table}[tbp]
\centering
\caption{Ticker codes of real data used in the experiment.}
\begin{tabular}{@{\extracolsep{\fill}}ccccccccc}
\toprule
2802 & 3382 & 4063 & 4452 & 4568 & 4578 & 6501 & 6502 & 7203\\
7267 & 8001 & 8035 & 8058 & 8306 & 8411 & 9202 & 9613 & 9984\\
\bottomrule
\end{tabular}
\label{Tab:tickers}
\end{table}

\begin{itemize}
\item Real data: We used FLEX-FULL historical tick data provided by the Japan Exchange Group~\citep{flex_full}. It includes order and execution series data, called tick data. By recording the mid price every minute, we obtained a one-minute bar price series with length $300$ per day. The data period was from January 5, 2015 to August 20, 2021. As described in Table~\ref{Tab:tickers}, we selected 18 stocks from those with sufficiently high liquidity during the period, ensuring a diverse range of industries. The number of stocks was $M=18$.
\item Number of time steps: In an artificial market simulation, a time step is defined as the period during which a randomly selected agent places orders. In our experiment, each simulation consisted of $T_{sim}=2,110$ time steps. For the first $100$ steps and from step $1,100$ to $1,110$, only order placement was allowed; no order execution occurred. We recorded every order and execution event with several statistics such as market/mid prices and order/execution volumes at each time step to form synthetic tick data.
\item Market: We set the number of artificial markets to $1$, tick size to $1.00\times10^{-4}$, initial market price to $p_0=300.00$, and fundamental price to a constant value of $\forall t~p_t^f=300.00$.
\item Agent: We set the value of following parameters regarding our agent model as follows with reference to \citet{fcn2}. $n=200$, $\lambda^f=10.00$, $\lambda^n=1.00$, $\sigma^n=0.01$, $\forall j~Pr[M_0^j=O]=0.5$, $\tau^{j,f}=200$, $\tau^j=100\times(1+w^{j,f})/(1+w^{j,c})$, $\alpha^j=\alpha\times(1+w^{j,f})/(1+w^{j,c})$, $\alpha\in\mathbb{R}_+$. All agents changed their mood at every time steps with probabilities described in Equation~(\ref{Eq:mood_transition}) sequentially in random order. 
\end{itemize}

We mainly utilized Python-based limit order book market simulator, PAMS~\citep{pams}, and OT library POT~\citep{pot} for our implementation.

\begin{table}[tbp]
\centering
\caption{Parameters to be calibrated and the candidate values.}
\begin{tabular}{ll}
\midrule
Parameter                   & Candidate values\\
\midrule
$c_0^j$                     & $U(0,30,000)$, $Pareto(5,000,1.50)$\\
$\lambda^c$                 & $0.00$, $1.50$, $1.75$, $2.00$, $2.25$, $2.50$\\
$\lambda^m~(\times10^{-3})$ & $0.00$, $0.01$, $0.02$, $0.03$, $0.04$, $0.05$\\
$\nu$                       & $0.30$, $0.50$, $0.70$\\
$\alpha$                    & $0.05$, $0.10$, $0.15$, $0.20$, $0.25$, $0.30$\\
\midrule
\end{tabular}
\label{Tab:searched_params}
\end{table}

\subsubsection{Parameters calibration}

In our experiment, we conducted parameter calibration as summarized in Table~\ref{Tab:searched_params}\footnote{$\lambda^c$ and $\lambda^m$ tended to destabilize simulations when they were set too high. We set the values of these variables in the range where the convergence rates of the numerical calculations by the agent exceeded $0.96$.}. For each of the $8$ simulations, we ran $100$ trials with different seed values for every possible parameter combination. We then recorded the parameter combination that yields the minimum average OT distance $\bar{OT}$. We fixed the values of parameters related to the candidate components causing the cubic law in simulations where they were not added (see Table~\ref{Tab:hypotheses_agents}). For example, in simulation $0$, since none of the three components were added to the agent model, the following parameters remained fixed: $c_0^j\sim U,~\lambda^c=0.00,~\lambda^m=0.00$.

We set $\beta=1.50$ following \citet{beta_1.5} and $c_{min}$ to $5,000$, ensuring that the expected value $\mathbb{E}[c_0^j]$ aligned with a uniform distribution $U(0,30,000)$.

\subsubsection{Assign Calendar Time to Synthetic Data}

\begin{figure}[b]
\centering
\includegraphics[bb=0 0 725 457, width=8.5cm]{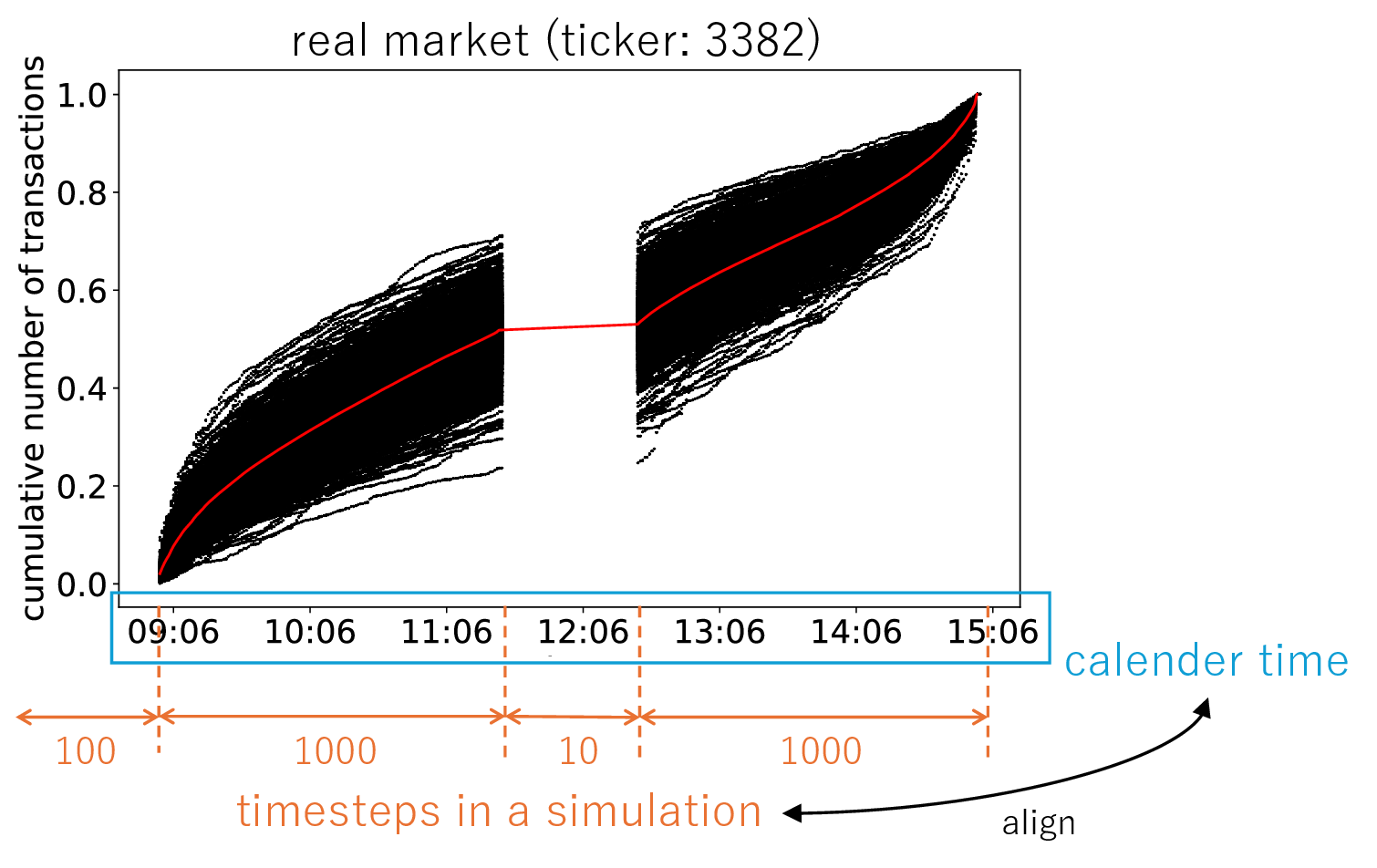}
\caption{Paths of the cumulative number of transactions scaled to 1 in real data. The ticker is $3382$. The red line represents the mean path. We randomly selected a single transactions path and assigned calendar time to synthetic tick data to ensure that the sampled path matched the transactions path of the resampled synthetic data.}
\label{Fig:transactions_time_series_3382}
\end{figure}

In an artificial market, a time step is defined as a process of submitting orders by an agent. Unlike real data, in which order data are arranged at unequal intervals, the time interval between when an investor submits an order is not considered. This gap needs to be filled to structure the simulation results and compare them with real intraday data. To achieve this objective, we assigned calendar time to synthetic data generated through simulations by the following procedure.

\begin{enumerate}
\item Record cumulative number (not volume) of transactions per minute from each intraday tick data. 
\item Scale the transactions path to have a final value of $1$. Figure~\ref{Fig:transactions_time_series_3382} presents the paths of scaled number of transactions in our real data.
\item Randomly sample the scaled number of transactions path.
\item Resample execution events in the artificial market simulations to ensure that the sampled real transactions path matches the transactions path of the synthetic data. For instance, if $2\%$ of the total number of daily transactions have occurred by 9:01 in the sampled real path, then assign a calendar time of 9:01 to the mid price at the time step corresponding to the $2\%$ cumulative number of transactions in the synthetic tick data.
\end{enumerate}

Using the above procedure, we converted the synthetic tick data into a one-minute bar price series with a structure identical to that of the real data. The number of data points $N$ was $461,700=300~\mathrm{(minutes~per~day)}\times 1,539~\mathrm{(days)}$ for each real data and $30,000=300~\mathrm{(minutes~per~trial)}\times 100~\mathrm{(trials)}$ for synthetic data. We set the sizes of real and synthetic point clouds $K=23,085$ and $L=1,500$. We uniformly sampled $1,500$ points from real point cloud $X_{real,m}$ in solving the OT problem for computational efficiency.

\subsection{Other Stylized Facts}\label{Sec:other_stylized_facts}
We verified that the synthetic data satisfied the following stylized facts by referring \citet{stylized_facts,get_real} to evaluate the validity of our experiment.

\begin{itemize}
\item The kurtosis of the stock return distribution $\kappa_r$ is positive.
\item The correlation between absolute return and execution volume $\rho=\mathrm{Corr}(|r_t|,v_t)$ is positive.
\item The auto-correlation of the absolute return series $\gamma(T)=\mathrm{Corr}(|r_t|,|r_{t-T}|)$ is positive.
\end{itemize}

\begin{table*}[tb]
\centering
\caption{The results of the simulations in Experiment 1. The symbols $(*)$ and $(+)$ denote parameters that are fixed and those that are searched according to Table~\ref{Tab:searched_params} in the simulations, respectively. $\bar{OT}$ represents the sample mean of the OT distances calculated between the synthetic point cloud $X_{syn}$ and the real point clouds $X_{real,m},~m=1,...,M$, and sample standard deviation is shown in parentheses. The Real row shows the average of the Hill indices of real data and the OT distances between real point clouds $OT(X_{real,m},X_{real,m'}),~ m,m'\in\{1,...,M\}$.}
\begin{tabular*}{16cm}{@{\extracolsep{\fill}}rrccccccc}
\toprule
 && \multicolumn{5}{c}{Params} & \multicolumn{2}{c}{Metrics}\\
  &  & $c_0^j$ & $\lambda^c$ & $\lambda^m~(\times10^{-3})$ & $\nu$ & $\alpha$ & $\hat{\zeta}_r$ & $\bar{OT}~(\times10^{-4})$\\ 
    \cmidrule(lr){3-7} \cmidrule(lr){8-9}
\multicolumn{2}{r}{Real} &$-$&$-$&$-$&$-$&$-$& $2.95$ & $8.57~(\pm6.55)$\\
\multirow{8}{*}{Simulation No.}&
  0 & $U$ $(*)$ & $0.00$ $(*)$ & $0.00$ $(*)$ & $0.00$ $(*)$ & $0.30$ $(+)$ & $4.03$ & $141.57~ (\pm33.59)$ \\
  &1 & $Pareto$ $(+)$ & $0.00$  $(*)$ & $0.00$ $(*)$ & $0.00$ $(*)$ & $0.25$  $(+)$ & $3.82$ & $88.26~ (\pm26.50)$ \\
  &2 & $U$ $(*)$ & $2.50$ $(+)$ & $0.00$ $(*)$ & $0.00$ $(*)$ & $0.25$ $(+)$ & $3.14$ & $13.36~ (\pm7.70)$ \\
  &3 & $U$ $(*)$ & $0.00$ $(*)$ & $0.05$ $(+)$ & $0.30$ $(+)$ & $0.30$ $(+)$ & $3.30$ & $25.45~ (\pm15.26)$ \\
  &4 & $Pareto$ $(+)$ & $2.00$ $(+)$ & $0.00$ $(*)$ & $0.00$ $(*)$ & $0.30$ $(+)$ & $3.04$ & $7.83~ (\pm5.87)$\\
  &5 & $Pareto$ $(+)$ & $0.00$ $(*)$ & $0.04$ $(+)$ & $0.50$ $(+)$ & $0.05$ $(+)$ & $3.53$ & $61.57~ (\pm24.87)$\\
  &6 & $U$ $(*)$ & $1.50$ $(+)$ & $0.01$ $(+)$ & $0.70$ $(+)$ & $0.05$ $(+)$ & $3.06$ & $9.31~ (\pm5.61)$\\
  &7 & $Pareto$ $(+)$ & $1.75$ $(+)$ & $0.04$ $(+)$ & $0.30$ $(+)$ & $0.20$ $(+)$ & $2.95$ & $6.06~ (\pm4.04)$\\
\midrule
\end{tabular*}
\label{Tab:sim_results}
\end{table*}

\section{Results and Discussion\label{Sec:results}}

\subsection{Magnitude of Influence by Each Component}
Table~\ref{Tab:sim_results} summarizes the parameter values selected through the calibrations for each simulation, along with the corresponding Hill indices and average OT distances to the real point clouds. A comparison of simulations $0$ with $1\sim3$, where each simulation introduced a different one of the three components, revealed that all components contributed to decreasing the Hill index and the average OT distance. The results of simulation $2$, in which the chartist trader was introduced, suggest that this component had the most significant impact. Similarly, in simulations $4$, $6$, and $7$, where the chartist trader was introduced, the resulting return distributions exhibited tails more closely resembling real data. This suggests that in a high-frequency domain, the informational effect of prices has the strongest impact on the power law of stock returns.

\begin{figure}[tbp]
\centering
\includegraphics[bb=0 0 546 226, width=8.5cm]{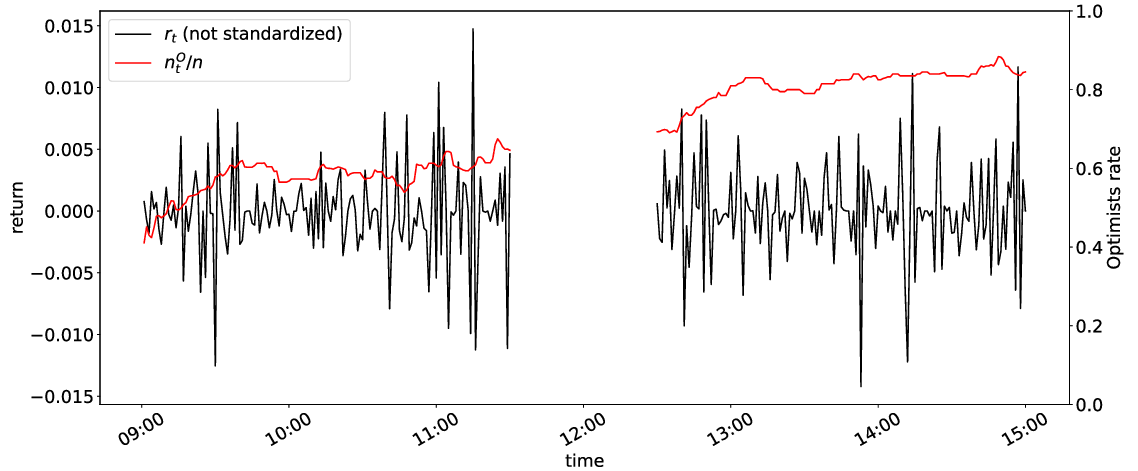}
\caption{The return and optimists' rate series in a representative trial of simulation $3$.}
\label{Fig:return_series_w_mood3}
\end{figure}

Conversely, the contribution of mood-aware traders was limited. In simulation $3$, the daily mood change rate, calculated as the following equation, had a sample mean of $0.24$ and a sample standard deviation of $0.07$ across $100$ trials.

\begin{eqnarray}
\max_{t\in\{1,...,T_{sim}\}}r_t^O-\min_{t'\in\{1,...,T_{sim}\}}r_{t'}^O
\end{eqnarray}
As shown in Figure~\ref{Fig:return_series_w_mood3}, simulation $3$ could not reproduce the phenomenon that the price fluctuates sharply as market mood undergoes a drastic change, probably because the simulation term was insufficient. Assuming that the market mood is cultivated over a timescale of more than a day, herd behavior is not dominant component for the tail of stock return distribution in high frequency domain.

\begin{table*}[tb]
\centering
\caption{Statistics regarding stylized facts calculated for real and synthetic data.}
\begin{tabular*}{14cm}{@{\extracolsep{\fill}}rrccccccc}
\toprule
                       &      & $\kappa_r$ & $\hat{\zeta}_r$ & $\rho$ & $\gamma(1)$ & $\gamma(10)$ & $\gamma(20)$ & $\gamma(30)$\\
\cmidrule(lr){3-9}
& Real  & 8.15       & 2.95            & 0.45   & 0.18        & 0.11         & 0.07        & 0.05         \\
\multirow{8}{*}{Simulation No.}   & 0    & 3.22       & 4.03            & 0.08   & 0.18        & 0.00         & -0.00       & 0.00         \\
                       & 1    & 3.96       & 3.82            & 0.03   & 0.20        & 0.00         & -0.00       & -0.00        \\
                       & 2    & 9.03       & 3.14            & 0.10   & 0.25        & 0.03         & 0.02        & 0.01         \\
                       & 3    & 5.65       & 3.30            & 0.06   & 0.16        & -0.00        & 0.00        & -0.00        \\
                       & 4    & 13.04      & 3.04            & 0.08   & 0.28        & 0.04         & 0.02        & 0.01         \\
                       & 5    & 4.53       & 3.53            & 0.06   & 0.27        & 0.01         & 0.01        & -0.00        \\
                       & 6    & 8.23       & 4.06            & 0.08   & 0.32        & 0.03         & 0.02        & 0.01         \\
                       & 7    & 11.18      & 2.95            & 0.07   & 0.31        & 0.08         & 0.05        & 0.04         \\
\bottomrule
\end{tabular*}
\label{Tab:other_stylized_facts}
\end{table*}

\subsection{Interaction Among The Components}

By assuming that the power law of demand size and the informational effect of prices act independently, we can estimate the theoretical Hill index for simulation $4$ based on the Hill indices obtained from simulations $0$, $1$, and $2$. Let $\hat{\zeta}_{r,s}$ denote the Hill index obtained from simulations $s~(s=0,1,2,4)$ when all parameters except for $c_0^j$ and $\lambda^c$ are fixed. The theoretical Hill index for simulation $4$, denoted by $\hat{\zeta}_{r,4}^{theoretical}$ is defined as follows.
\begin{eqnarray}
\hat{\zeta}_{r,4}^{theoretical}=\hat{\zeta}_{r,0}-(\hat{\zeta}_{r,0}-\hat{\zeta}_{r,1})-(\hat{\zeta}_{r,0}-\hat{\zeta}_{r,2})\label{Eq:theoretical_zeta}
\end{eqnarray}

\begin{figure}[tbp]
\centering
\includegraphics[bb=0 0 512 218, width=8.5cm]{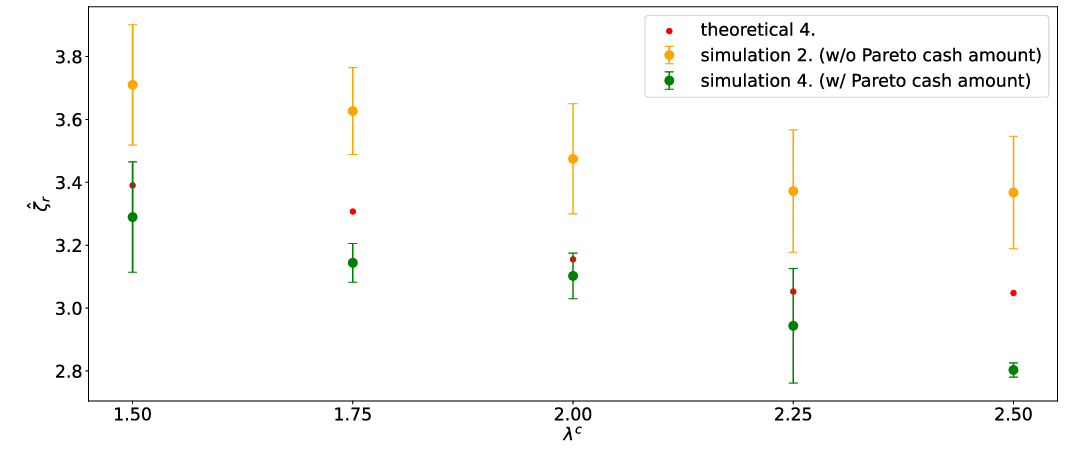}
\caption{The relationship between $\lambda^c$ and $\hat{\zeta}_r$ in simulations $2$ and $4$. The dots with the error bars represent the sample mean and standard deviation for all combinations of parameters searched, as detailed in Table~\ref{Tab:searched_params}. The red dots represent the sample mean of the theoretically calculated $\hat{\zeta}_{r,4}^{theoretical}$ based on the results of simulations $1$ and $2$, obtained from all parameter combinations.}
\label{Fig:hill_index_w_different_lambdac}
\end{figure}

Figure~\ref{Fig:hill_index_w_different_lambdac} compares the theoretical Hill index $\hat{\zeta}_{r,4}^{theoretical}$ with the observed Hill index $\hat{\zeta}_{r,4}$ from simulation $4$ for various values of $\lambda^c$. The figure reveals that when both the power law of demand size and informational effect of prices are incorporated in to the agent model, the Hill index decreases more significantly than the simple sum of the individual effects, as suggested by Equation~(\ref{Eq:theoretical_zeta}).

This finding of a synergistic relationship between the the power law of demand size and informational effect of prices implies the significance of indirect market impact. Market impact refers to the fluctuation in prices in financial markets owing to orders exceeding market liquidity. Large orders cause not only direct but also what we call indirect market impacts, in which large orders provide additional information to other market participants and cause secondary fluctuations. Our analysis suggests that the coexistence of the two components can generate indirect market impact and contribute to the universal cubic law. For example, the Flash Crash in the U.S. stock market on May 6, 2010, at which a large sell order was said to be followed by algorithmic sell orders and market prices dropped sharply~\citep{flash_crash}, can be understood as the direct and indirect market impact phenomenon. 

Given the observed synergistic effect, it is essential to implement mitigation strategies that address both sides of the interaction in a complementary manner. For example, on the one hand, order splitting guidelines can reduce the direct market impact of large trades. On the other hand, alternative markets such as dark pool markets can help control the dissemination of order-related information~\citep{darkpool}, thereby mitigating indirect market impact.

\subsection{Other Stylized Facts}

Table~\ref{Tab:other_stylized_facts} summarizes the descriptive statistics calculated for real and synthetic data used in our experiments. The table demonstrates two key points: 1)the validity of our experiments by showing that the synthetic data meets stylized facts, and 2)the extensibility of our methodology to examine components beyond the cubic law. Regarding 1), Simulations 2, 4, 6, and 7 satisfy all the conditions listed in Section~\ref{Sec:other_stylized_facts}. Simulations 0, 1, 3, and 5 were run without the chartist trader, which is an important component that contributes to the stylized facts, as also discussed in \citet{fcn2}. However, regarding 2), in our synthetic data, the values of $\rho$, $\gamma(10)$, $\gamma(20)$, and $\gamma(30)$ were not remarkable as those in real data, although the values meet the standards established by prior research for validating simulations. These results suggest the existence of components not modeled in our study such as behavioral switching~\citep{behavioral_switching} or investor inertia~\citep{investor_inertia} that contribute to these stylized facts~\citep{volalitily_clustering_mechanism, volume_volatility_correlation_mechanism}.

\section{Conclusion}
We proposed a method to factorize the underlying components responsible for financial stylized facts using artificial market simulations with quantitative evaluation using OT, focusing on the universal cubic law as an example---a power law distribution of stock returns with an exponent of 3---and experimented using the proposed method. Drawing from previous research, we considered three candidate components and placed each candidate on our agent model. Our experiment demonstrated that all components contribute to weighting the tail of the return distribution and that the informational effect of prices plays a major role in generating the power law distribution of returns in the high-frequency domain. We also confirmed that the effect of multiple components is greater than the simple sum of their individual effects. This synergistic effect could only be uncovered through controlled experiments using artificial markets, in combination with the robust quantitative evaluation enabled by OT. 

Our future work can be directed towards two main areas: 1)conducting long-term simulations, such as simulations over the days to determine whether the primary components influencing the cubic law vary across these time domains, and 2)applying our proposed methodology to other stylized facts such as volume-volatility correlation and volatility clustering to make artificial market realistic and interpretable in other aspects.


{\small
\bibliographystyle{ACM-Reference-Format}
\bibliography{sample-base}}


\begin{thebibliography}{41}


\ifx \showCODEN    \undefined \def \showCODEN     #1{\unskip}     \fi
\ifx \showISBNx    \undefined \def \showISBNx     #1{\unskip}     \fi
\ifx \showISBNxiii \undefined \def \showISBNxiii  #1{\unskip}     \fi
\ifx \showISSN     \undefined \def \showISSN      #1{\unskip}     \fi
\ifx \showLCCN     \undefined \def \showLCCN      #1{\unskip}     \fi
\ifx \shownote     \undefined \def \shownote      #1{#1}          \fi
\ifx \showarticletitle \undefined \def \showarticletitle #1{#1}   \fi
\ifx \showURL      \undefined \def \showURL       {\relax}        \fi
\providecommand\bibfield[2]{#2}
\providecommand\bibinfo[2]{#2}
\providecommand\natexlab[1]{#1}
\providecommand\showeprint[2][]{arXiv:#2}

\bibitem[Alfarano and Lux(2007)]%
        {herd_behavior}
\bibfield{author}{\bibinfo{person}{Simone Alfarano} {and}
  \bibinfo{person}{Thomas Lux}.} \bibinfo{year}{2007}\natexlab{}.
\newblock \showarticletitle{A noise trader model as a generator of apparent
  financial power laws and long memory}.
\newblock \bibinfo{journal}{\emph{Macroeconomic Dynamics}}
  \bibinfo{volume}{11}, \bibinfo{number}{S1} (\bibinfo{year}{2007}),
  \bibinfo{pages}{80--101}.
\newblock


\bibitem[Artzner et~al\mbox{.}(1999)]%
        {var_pitfalls}
\bibfield{author}{\bibinfo{person}{Philippe Artzner}, \bibinfo{person}{Freddy
  Delbaen}, \bibinfo{person}{Jean-Marc Eber}, {and} \bibinfo{person}{David
  Heath}.} \bibinfo{year}{1999}\natexlab{}.
\newblock \showarticletitle{Coherent measures of risk}.
\newblock \bibinfo{journal}{\emph{Mathematical Finance}} \bibinfo{volume}{9},
  \bibinfo{number}{3} (\bibinfo{year}{1999}), \bibinfo{pages}{203--228}.
\newblock


\bibitem[Assefa et~al\mbox{.}(2021)]%
        {financial_synthetic_data_generation1}
\bibfield{author}{\bibinfo{person}{Samuel~A. Assefa}, \bibinfo{person}{Danial
  Dervovic}, \bibinfo{person}{Mahmoud Mahfouz}, \bibinfo{person}{Robert~E.
  Tillman}, \bibinfo{person}{Prashant Reddy}, {and} \bibinfo{person}{Manuela
  Veloso}.} \bibinfo{year}{2021}\natexlab{}.
\newblock \showarticletitle{{Generating synthetic data in finance:
  Opportunities, challenges and pitfalls}}. In
  \bibinfo{booktitle}{\emph{Proceedings of the ACM International Conference on
  AI in Finance}}. Article \bibinfo{articleno}{44},
  \bibinfo{numpages}{8}~pages.
\newblock


\bibitem[Battiston et~al\mbox{.}(2016)]%
        {am3}
\bibfield{author}{\bibinfo{person}{Stefano Battiston},
  \bibinfo{person}{J.~Doyne Farmer}, \bibinfo{person}{Andreas Flache},
  \bibinfo{person}{Diego Garlaschelli}, \bibinfo{person}{Andrew~G. Haldane},
  \bibinfo{person}{Hans Heesterbeek}, \bibinfo{person}{Cars Hommes},
  \bibinfo{person}{Carlo Jaeger}, \bibinfo{person}{Robert May}, {and}
  \bibinfo{person}{Marten Scheffer}.} \bibinfo{year}{2016}\natexlab{}.
\newblock \showarticletitle{Complexity theory and financial regulation}.
\newblock \bibinfo{journal}{\emph{Science}} \bibinfo{volume}{351},
  \bibinfo{number}{6275} (\bibinfo{year}{2016}), \bibinfo{pages}{818--819}.
\newblock


\bibitem[Bayraktar et~al\mbox{.}(2006)]%
        {investor_inertia}
\bibfield{author}{\bibinfo{person}{Erhan Bayraktar}, \bibinfo{person}{Ulrich
  Horst}, {and} \bibinfo{person}{Ronnie Sircar}.}
  \bibinfo{year}{2006}\natexlab{}.
\newblock \showarticletitle{A limit theorem for financial markets with inert
  investors}.
\newblock \bibinfo{journal}{\emph{Mathematics of Operations Research}}
  \bibinfo{volume}{31}, \bibinfo{number}{4} (\bibinfo{year}{2006}),
  \bibinfo{pages}{789--810}.
\newblock


\bibitem[Black(1986)]%
        {noise}
\bibfield{author}{\bibinfo{person}{Fischer Black}.}
  \bibinfo{year}{1986}\natexlab{}.
\newblock \showarticletitle{Noise}.
\newblock \bibinfo{journal}{\emph{The Journal of Finance}}
  \bibinfo{volume}{41}, \bibinfo{number}{3} (\bibinfo{year}{1986}),
  \bibinfo{pages}{528--543}.
\newblock


\bibitem[Bookstaber(2017)]%
        {am2}
\bibfield{author}{\bibinfo{person}{Richard Bookstaber}.}
  \bibinfo{year}{2017}\natexlab{}.
\newblock \showarticletitle{Agent-based models for financial crises}.
\newblock \bibinfo{journal}{\emph{Annual Review of Financial Economics}}
  \bibinfo{volume}{9} (\bibinfo{year}{2017}), \bibinfo{pages}{85--100}.
\newblock


\bibitem[Chiarella and Iori(2002)]%
        {fcn1}
\bibfield{author}{\bibinfo{person}{Carl Chiarella} {and}
  \bibinfo{person}{Giulia Iori}.} \bibinfo{year}{2002}\natexlab{}.
\newblock \showarticletitle{A simulation analysis of the microstructure of
  double auction markets}.
\newblock \bibinfo{journal}{\emph{Quantitative Finance}} \bibinfo{volume}{2},
  \bibinfo{number}{5} (\bibinfo{year}{2002}), \bibinfo{pages}{346--353}.
\newblock


\bibitem[Chiarella et~al\mbox{.}(2009)]%
        {fcn2}
\bibfield{author}{\bibinfo{person}{Carl Chiarella}, \bibinfo{person}{Giulia
  Iori}, {and} \bibinfo{person}{Josep Perell\'{\o}}.}
  \bibinfo{year}{2009}\natexlab{}.
\newblock \showarticletitle{The impact of heterogeneous trading rules on the
  limit order book and order flows}.
\newblock \bibinfo{journal}{\emph{Journal of Economic Dynamics and Control}}
  \bibinfo{volume}{33}, \bibinfo{number}{3} (\bibinfo{year}{2009}),
  \bibinfo{pages}{525--537}.
\newblock


\bibitem[Cont(2007)]%
        {volalitily_clustering_mechanism}
\bibfield{author}{\bibinfo{person}{Rama Cont}.}
  \bibinfo{year}{2007}\natexlab{}.
\newblock \bibinfo{booktitle}{\emph{Volatility clustering in financial markets:
  Empirical facts and agent-based models}}.
\newblock \bibinfo{address}{Berlin, Heidelberg}, \bibinfo{pages}{289--309}.
\newblock


\bibitem[Daniel(1989)]%
        {msm1}
\bibfield{author}{\bibinfo{person}{McFadden Daniel}.}
  \bibinfo{year}{1989}\natexlab{}.
\newblock \showarticletitle{A method of simulated moments for estimation of
  discrete response models without numerical integration}.
\newblock \bibinfo{journal}{\emph{Econometrica}} \bibinfo{volume}{57},
  \bibinfo{number}{5} (\bibinfo{year}{1989}), \bibinfo{pages}{995--1026}.
\newblock


\bibitem[Domowitz et~al\mbox{.}(2008)]%
        {darkpool}
\bibfield{author}{\bibinfo{person}{Ian Domowitz}, \bibinfo{person}{Ilya
  Finkelshteyn}, {and} \bibinfo{person}{Henry Yegerman}.}
  \bibinfo{year}{2008}\natexlab{}.
\newblock \showarticletitle{{Cul de sacs and highways: An optical tour of dark
  pool trading performance}}.
\newblock \bibinfo{journal}{\emph{The Journal Of Trading}}  \bibinfo{volume}{4}
  (\bibinfo{year}{2008}), \bibinfo{pages}{16--22}.
\newblock


\bibitem[Fagereng et~al\mbox{.}(2020)]%
        {rich_get_richer_return}
\bibfield{author}{\bibinfo{person}{Andreas Fagereng}, \bibinfo{person}{Luigi
  Guiso}, \bibinfo{person}{Davide Malacrino}, {and} \bibinfo{person}{Luigi
  Pistaferri}.} \bibinfo{year}{2020}\natexlab{}.
\newblock \showarticletitle{Heterogeneity and persistence in returns to
  wealth}.
\newblock \bibinfo{journal}{\emph{Econometrica}} \bibinfo{volume}{88},
  \bibinfo{number}{1} (\bibinfo{year}{2020}), \bibinfo{pages}{115--170}.
\newblock


\bibitem[Farmer and Foley(2009)]%
        {am1}
\bibfield{author}{\bibinfo{person}{J.~Doyne Farmer} {and}
  \bibinfo{person}{Duncan Foley}.} \bibinfo{year}{2009}\natexlab{}.
\newblock \showarticletitle{The economy needs agent-based modeling}.
\newblock \bibinfo{journal}{\emph{Nature}}  \bibinfo{volume}{460}
  (\bibinfo{year}{2009}), \bibinfo{pages}{685--6}.
\newblock


\bibitem[Flamary et~al\mbox{.}(2021)]%
        {pot}
\bibfield{author}{\bibinfo{person}{R{\'e}mi Flamary}, \bibinfo{person}{Nicolas
  Courty}, \bibinfo{person}{Alexandre Gramfort}, \bibinfo{person}{Mokhtar~Z.
  Alaya}, \bibinfo{person}{Aur{\'e}lie Boisbunon}, \bibinfo{person}{Stanislas
  Chambon}, \bibinfo{person}{Laetitia Chapel}, \bibinfo{person}{Adrien
  Corenflos}, \bibinfo{person}{Kilian Fatras}, \bibinfo{person}{Nemo Fournier},
  \bibinfo{person}{L{\'e}o Gautheron}, \bibinfo{person}{Nathalie~T.H. Gayraud},
  \bibinfo{person}{Hicham Janati}, \bibinfo{person}{Alain Rakotomamonjy},
  \bibinfo{person}{Ievgen Redko}, \bibinfo{person}{Antoine Rolet},
  \bibinfo{person}{Antony Schutz}, \bibinfo{person}{Vivien Seguy},
  \bibinfo{person}{Danica~J. Sutherland}, \bibinfo{person}{Romain Tavenard},
  \bibinfo{person}{Alexander Tong}, {and} \bibinfo{person}{Titouan Vayer}.}
  \bibinfo{year}{2021}\natexlab{}.
\newblock \showarticletitle{{POT: Python optimal transport}}.
\newblock \bibinfo{journal}{\emph{Journal of Machine Learning Research}}
  \bibinfo{volume}{22}, \bibinfo{number}{78} (\bibinfo{year}{2021}),
  \bibinfo{pages}{1--8}.
\newblock


\bibitem[Franke(2009)]%
        {msm2}
\bibfield{author}{\bibinfo{person}{Reiner Franke}.}
  \bibinfo{year}{2009}\natexlab{}.
\newblock \showarticletitle{Applying the method of simulated moments to
  estimate a small agent-based asset pricing model}.
\newblock \bibinfo{journal}{\emph{Journal of Empirical Finance}}
  \bibinfo{volume}{16}, \bibinfo{number}{5} (\bibinfo{year}{2009}),
  \bibinfo{pages}{804--815}.
\newblock


\bibitem[Gabaix et~al\mbox{.}(2003)]%
        {empirical_power_law_return1}
\bibfield{author}{\bibinfo{person}{Xavier Gabaix},
  \bibinfo{person}{Parameswaran Gopikrishnan}, \bibinfo{person}{Vasiliki
  Plerou}, {and} \bibinfo{person}{H. Stanley}.}
  \bibinfo{year}{2003}\natexlab{}.
\newblock \showarticletitle{A theory of power-law distributions in financial
  market fluctuations}.
\newblock \bibinfo{journal}{\emph{Nature}}  \bibinfo{volume}{423}
  (\bibinfo{year}{2003}), \bibinfo{pages}{267--70}.
\newblock


\bibitem[Gabaix et~al\mbox{.}(2006)]%
        {cubic_law}
\bibfield{author}{\bibinfo{person}{Xavier Gabaix},
  \bibinfo{person}{Parameswaran Gopikrishnan}, \bibinfo{person}{Vasiliki
  Plerou}, {and} \bibinfo{person}{H.~Eugene Stanley}.}
  \bibinfo{year}{2006}\natexlab{}.
\newblock \showarticletitle{Institutional investors and stock market
  volatility}.
\newblock \bibinfo{journal}{\emph{Quarterly Journal of Economics}}
  \bibinfo{volume}{121}, \bibinfo{number}{2} (\bibinfo{year}{2006}),
  \bibinfo{pages}{461--504}.
\newblock


\bibitem[Gabriel and Marco(2019)]%
        {ot}
\bibfield{author}{\bibinfo{person}{Peyr^^c3^^a9 Gabriel} {and}
  \bibinfo{person}{Cuturi Marco}.} \bibinfo{year}{2019}\natexlab{}.
\newblock \showarticletitle{{Computational optimal transport: With applications
  to data science}}.
\newblock \bibinfo{journal}{\emph{Foundations and Trends^^c2^^ae in Machine
  Learning}} \bibinfo{volume}{11}, \bibinfo{number}{5-6}
  (\bibinfo{year}{2019}), \bibinfo{pages}{355--607}.
\newblock


\bibitem[Gao et~al\mbox{.}(2022)]%
        {stylized_facts_distance}
\bibfield{author}{\bibinfo{person}{Kang Gao}, \bibinfo{person}{Perukrishnen
  Vytelingum}, \bibinfo{person}{Stephen Weston}, \bibinfo{person}{Wayne Luk},
  {and} \bibinfo{person}{Ce Guo}.} \bibinfo{year}{2022}\natexlab{}.
\newblock \bibinfo{title}{{Understanding intra-day price formation process by
  agent-based financial market simulation: Calibrating the extended chiarella
  model}}.
\newblock
\showeprint[arxiv]{2208.14207}


\bibitem[Gao et~al\mbox{.}(2024)]%
        {am4}
\bibfield{author}{\bibinfo{person}{Kang Gao}, \bibinfo{person}{Perukrishnen
  Vytelingum}, \bibinfo{person}{Stephen Weston}, \bibinfo{person}{Wayne Luk},
  {and} \bibinfo{person}{Ce Guo}.} \bibinfo{year}{2024}\natexlab{}.
\newblock \showarticletitle{{High-frequency financial market simulation and
  flash crash scenarios analysis: An agent-based modelling approach}}.
\newblock \bibinfo{journal}{\emph{Journal of Artificial Societies and Social
  Simulation}} \bibinfo{volume}{27}, \bibinfo{number}{2}
  (\bibinfo{year}{2024}).
\newblock


\bibitem[Glielmo et~al\mbox{.}(2023)]%
        {msm_rl}
\bibfield{author}{\bibinfo{person}{Aldo Glielmo}, \bibinfo{person}{Marco
  Favorito}, \bibinfo{person}{Debmallya Chanda}, {and}
  \bibinfo{person}{Domenico Delli~Gatti}.} \bibinfo{year}{2023}\natexlab{}.
\newblock \showarticletitle{Reinforcement learning for combining search methods
  in the calibration of economic ABMs}. In
  \bibinfo{booktitle}{\emph{Proceedings of the ACM International Conference on
  AI in Finance}}. \bibinfo{pages}{305--313}.
\newblock


\bibitem[Gopikrishnan et~al\mbox{.}(1999)]%
        {empirical_power_law_return3}
\bibfield{author}{\bibinfo{person}{Parameswaran Gopikrishnan},
  \bibinfo{person}{Vasiliki Plerou}, \bibinfo{person}{Lu\'{\i}s~A.
  Nunes~Amaral}, \bibinfo{person}{Martin Meyer}, {and}
  \bibinfo{person}{H.~Eugene Stanley}.} \bibinfo{year}{1999}\natexlab{}.
\newblock \showarticletitle{Scaling of the distribution of fluctuations of
  financial market indices}.
\newblock \bibinfo{journal}{\emph{Physical review. E, Statistical physics,
  plasmas, fluids, and related interdisciplinary topics}}  \bibinfo{volume}{60}
  (\bibinfo{year}{1999}), \bibinfo{pages}{5305--16}.
\newblock
Issue 5.


\bibitem[Hill(1975)]%
        {hill}
\bibfield{author}{\bibinfo{person}{Bruce~M. Hill}.}
  \bibinfo{year}{1975}\natexlab{}.
\newblock \showarticletitle{A simple general approach to inference about the
  tail of a distribution}.
\newblock \bibinfo{journal}{\emph{The Annals of Statistics}}
  \bibinfo{volume}{3}, \bibinfo{number}{5} (\bibinfo{year}{1975}),
  \bibinfo{pages}{1163--1174}.
\newblock


\bibitem[Hirano et~al\mbox{.}(2022)]%
        {lstm_hftmm}
\bibfield{author}{\bibinfo{person}{Masanori Hirano}, \bibinfo{person}{Kiyoshi
  Izumi}, {and} \bibinfo{person}{Hiroki Sakaji}.}
  \bibinfo{year}{2022}\natexlab{}.
\newblock \showarticletitle{Implementation of actual data for artificial market
  simulation}. In \bibinfo{booktitle}{\emph{The International Conference on
  Autonomous Agents and Multiagent Systems}}. \bibinfo{pages}{1624--1626}.
\newblock


\bibitem[Hirano et~al\mbox{.}(2023)]%
        {pams}
\bibfield{author}{\bibinfo{person}{Masanori Hirano}, \bibinfo{person}{Ryosuke
  Takata}, {and} \bibinfo{person}{Kiyoshi Izumi}.}
  \bibinfo{year}{2023}\natexlab{}.
\newblock \bibinfo{title}{{PAMS: Platform for artificial market simulations}}.
\newblock
\showeprint[arxiv]{2309.10729}


\bibitem[{Japan Exchange Group}(2025)]%
        {flex_full}
\bibfield{author}{\bibinfo{person}{{Japan Exchange Group}}.}
  \bibinfo{year}{2025}\natexlab{}.
\newblock \bibinfo{title}{Historical Data}.
\newblock
  \bibinfo{howpublished}{\url{https://www.jpx.co.jp/english/markets/paid-info-equities/historical/01.html}}.
\newblock


\bibitem[Karpoff(1987)]%
        {volume_volatility_correlation_mechanism}
\bibfield{author}{\bibinfo{person}{Jonathan~M. Karpoff}.}
  \bibinfo{year}{1987}\natexlab{}.
\newblock \showarticletitle{{The relation between price changes and trading
  volume: A survey}}.
\newblock \bibinfo{journal}{\emph{The Journal of Financial and Quantitative
  Analysis}} \bibinfo{volume}{22}, \bibinfo{number}{1} (\bibinfo{year}{1987}),
  \bibinfo{pages}{109--126}.
\newblock


\bibitem[Kirilenko et~al\mbox{.}(2017)]%
        {flash_crash}
\bibfield{author}{\bibinfo{person}{Andrei Kirilenko},
  \bibinfo{person}{Albert~S. Kyle}, \bibinfo{person}{Mehrdaa Samadi}, {and}
  \bibinfo{person}{Tugkan Tuzun}.} \bibinfo{year}{2017}\natexlab{}.
\newblock \showarticletitle{{The flash crash: High-frequency trading in an
  electronic market}}.
\newblock \bibinfo{journal}{\emph{The Journal of Finance}}
  \bibinfo{volume}{72}, \bibinfo{number}{3} (\bibinfo{year}{2017}),
  \bibinfo{pages}{967--998}.
\newblock


\bibitem[Lux(1998)]%
        {empirical_power_law_return2}
\bibfield{author}{\bibinfo{person}{Thomas Lux}.}
  \bibinfo{year}{1998}\natexlab{}.
\newblock \showarticletitle{{The limiting extremal behaviour of speculative
  returns: An analysis of intra-daily data from the frankfurt stock exchange}}.
\newblock \bibinfo{journal}{\emph{IFAC Proceedings Volumes}}
  \bibinfo{volume}{31}, \bibinfo{number}{16} (\bibinfo{year}{1998}),
  \bibinfo{pages}{29--33}.
\newblock


\bibitem[Lux and Alfarano(2016)]%
        {financial_power_laws_survey}
\bibfield{author}{\bibinfo{person}{Thomas Lux} {and} \bibinfo{person}{Simone
  Alfarano}.} \bibinfo{year}{2016}\natexlab{}.
\newblock \showarticletitle{{Financial power laws: Empirical evidence, models,
  and mechanisms}}.
\newblock \bibinfo{journal}{\emph{Chaos, Solitons $\&$ Fractals}}
  \bibinfo{volume}{88} (\bibinfo{year}{2016}), \bibinfo{pages}{3--18}.
\newblock


\bibitem[Lux and Marchesi(1998)]%
        {behavioral_switching}
\bibfield{author}{\bibinfo{person}{Thomas Lux} {and} \bibinfo{person}{Michele
  Marchesi}.} \bibinfo{year}{1998}\natexlab{}.
\newblock \showarticletitle{{Volatility clustering in financial markets: A
  micro-simulation of interacting agents}}.
\newblock \bibinfo{journal}{\emph{IFAC Proceedings Volumes}}
  \bibinfo{volume}{31}, \bibinfo{number}{16} (\bibinfo{year}{1998}),
  \bibinfo{pages}{7--10}.
\newblock


\bibitem[Lux and Marchesi(1999)]%
        {mas_essential}
\bibfield{author}{\bibinfo{person}{Thomas Lux} {and} \bibinfo{person}{Michele
  Marchesi}.} \bibinfo{year}{1999}\natexlab{}.
\newblock \showarticletitle{Scaling and criticality in a stochastic multi-agent
  model of a financial market}.
\newblock \bibinfo{journal}{\emph{Nature}}  \bibinfo{volume}{397}
  (\bibinfo{year}{1999}), \bibinfo{pages}{498--500}.
\newblock


\bibitem[Mantegna and Stanley(1995)]%
        {power_law_return}
\bibfield{author}{\bibinfo{person}{Rosario~N. Mantegna} {and}
  \bibinfo{person}{H.~Eugene Stanley}.} \bibinfo{year}{1995}\natexlab{}.
\newblock \showarticletitle{Scaling behavior in the dynamics of an economic
  index}.
\newblock \bibinfo{journal}{\emph{Nature}}  \bibinfo{volume}{376}
  (\bibinfo{year}{1995}), \bibinfo{pages}{46--49}.
\newblock


\bibitem[Nirei et~al\mbox{.}(2020)]%
        {info_effect}
\bibfield{author}{\bibinfo{person}{Makoto Nirei}, \bibinfo{person}{John
  Stachurski}, {and} \bibinfo{person}{Tsutomu Watanabe}.}
  \bibinfo{year}{2020}\natexlab{}.
\newblock \showarticletitle{Trade clustering and power laws in financial
  markets}.
\newblock \bibinfo{journal}{\emph{Theoretical Economics}} \bibinfo{volume}{15},
  \bibinfo{number}{4} (\bibinfo{year}{2020}), \bibinfo{pages}{1365--1398}.
\newblock


\bibitem[Rama(2001)]%
        {stylized_facts}
\bibfield{author}{\bibinfo{person}{Cont Rama}.}
  \bibinfo{year}{2001}\natexlab{}.
\newblock \showarticletitle{{Empirical properties of asset returns: Stylized
  facts and statistical issues}}.
\newblock \bibinfo{journal}{\emph{Quantitative Finance}} \bibinfo{volume}{1},
  \bibinfo{number}{2} (\bibinfo{year}{2001}), \bibinfo{pages}{223--236}.
\newblock


\bibitem[Santambrogio(2010)]%
        {ot2}
\bibfield{author}{\bibinfo{person}{Filippo Santambrogio}.}
  \bibinfo{year}{2010}\natexlab{}.
\newblock \bibinfo{title}{Introduction to optimal transport theory}.
\newblock
\showeprint[arxiv]{1009.3856}


\bibitem[Shiller(1984)]%
        {feedback}
\bibfield{author}{\bibinfo{person}{Robert~J. Shiller}.}
  \bibinfo{year}{1984}\natexlab{}.
\newblock \showarticletitle{Stock prices and social dynamics}.
\newblock \bibinfo{journal}{\emph{Brookings Papers on Economic Activity}}
  \bibinfo{volume}{15}, \bibinfo{number}{2} (\bibinfo{year}{1984}),
  \bibinfo{pages}{457--510}.
\newblock


\bibitem[Solomon and Richmond(2001)]%
        {beta_1.5}
\bibfield{author}{\bibinfo{person}{Sorin Solomon} {and} \bibinfo{person}{Peter
  Richmond}.} \bibinfo{year}{2001}\natexlab{}.
\newblock \showarticletitle{Power laws of wealth, market order volumes and
  market returns}.
\newblock \bibinfo{journal}{\emph{Physica A: Statistical Mechanics and its
  Applications}} \bibinfo{volume}{299}, \bibinfo{number}{1}
  (\bibinfo{year}{2001}), \bibinfo{pages}{188--197}.
\newblock


\bibitem[Stillman et~al\mbox{.}(2023)]%
        {calibration_w_nde}
\bibfield{author}{\bibinfo{person}{Namid~R Stillman}, \bibinfo{person}{Rory
  Baggott}, \bibinfo{person}{Justin Lyon}, \bibinfo{person}{Jianfei Zhang},
  \bibinfo{person}{Dingqui Zhu}, \bibinfo{person}{Tao Chen}, {and}
  \bibinfo{person}{Perukrishnen Vytelingum}.} \bibinfo{year}{2023}\natexlab{}.
\newblock \showarticletitle{Deep calibration of market simulations using neural
  density estimators and embedding networks}. In
  \bibinfo{booktitle}{\emph{Proceedings of the ACM International Conference on
  AI in Finance}}. \bibinfo{pages}{46--54}.
\newblock


\bibitem[Vyetrenko et~al\mbox{.}(2021)]%
        {get_real}
\bibfield{author}{\bibinfo{person}{Svitlana Vyetrenko}, \bibinfo{person}{David
  Byrd}, \bibinfo{person}{Nick Petosa}, \bibinfo{person}{Mahmoud Mahfouz},
  \bibinfo{person}{Danial Dervovic}, \bibinfo{person}{Manuela Veloso}, {and}
  \bibinfo{person}{Tucker Balch}.} \bibinfo{year}{2021}\natexlab{}.
\newblock \showarticletitle{{Get real: Realism metrics for robust limit order
  book market simulations}}. In \bibinfo{booktitle}{\emph{Proceedings of the
  ACM International Conference on AI in Finance}}. Article
  \bibinfo{articleno}{2}, \bibinfo{numpages}{8}~pages.
\newblock


\end{thebibliography}



\end{document}